 \documentclass{aa}  
 \usepackage{graphicx}
 \usepackage{txfonts}
\begin{document}

 \newcommand{\vdag}{(v)^\dagger}
 \newcommand{\oneskip}{\vskip\baselineskip}
 \newcommand{\cc}{\hbox{$\mu$}}
 \newcommand{\ergsec}{\hbox{erg s$^{-1}$}}
 \newcommand{\bb}{\hbox{$\nu$}}
 \newcommand{\dy}{\hbox{d$^{-1}$}}
 \newcommand{\so}{V2246 Cyg}
 \newcommand{\sx}{EXO 2030+375 }

 \def\hr{\hbox{$^{\rm h}$}}                 
 \def\fhr{\hbox{$.\!\!^{\rm h}$}}           
 \def\deg{\hbox{$^\circ$}}                  
 \def\fdeg{\hbox{$.\!\!^\circ$}}            
 \def\sec{\hbox{$^{\rm s}$}}                
 \def\fsec{\hbox{$.\!\!^{\rm s}$}}          
 \def\arcm{\hbox{$^\prime$}}                
 \def\farcm{\hbox{$.\mkern-4mu^\prime$}}    
 \def\arcs{\hbox{$^{\prime\prime}$}}        
 \def\farcs{\hbox{$.\!\!^{\prime\prime}$}}  
 \def\day{\hbox{$^{\rm d}$}}                
 \def\fday{\hbox{$.\!\!^{\rm d}$}}          
 \def\per{\hbox{$^{\rm p}$}}                
 \def\fper{\hbox{$.\!\!^{\rm p}$}}          
 \def\mm{\hbox{$^{\rm m}$}}                 
 \def\fmm{\hbox{$.\!\!^{\rm m}$}}           


 \title{Recent RXTE/ASM and ROTSEIIId observations of \\ EXO 2030+375 (\so~) }
 \authorrunning{Baykal et al. }
 \author{A. Baykal,
        \"U. K{\i}z{\i}lo\u{g}lu,
        N. K{\i}z{\i}lo\u{g}lu,
        E. Beklen,
        M. \"Ozbey 
        } 


  \institute{Physics Department, Middle East Technical University, Ankara 06531, Turkey
 }

 \date{}

\abstract
 {}
{ Using the archival RXTE/ASM and SWIFT/BAT observations,
the new orbital phases of Type I outbursts of EXO 2030+375 are
estimated. A possible correlation between the Type II outburst
and optical brightness variations is  investigated. 
    } 
{In order to estimate the phases of Type I outbursts,
we fitted Gaussian profiles to the RXTE/ASM and SWIFT/BAT light curves.
The time corresponding to the maximum value of the profiles is
treated as the arrival time of Type I outburst.  We used differential magnitudes 
in the time-series analysis of the optical light curve.
MIDAS and its suitable packages were used to reduce and analyze the spectra.
    }
{
Prior to the Type II outburst, orbital phases of Type I outbursts were 
delayed for $\sim$ 6 days after the periastron passage, which is consistent with 
findings of Wilson et al. (2002, 2005).  After the giant Type II outburst, 
the phase of Type I outbursts underwent a sudden shift of $\sim$13 days after 
the periastron passage. The amplitudes of Type I outbursts
were increased  between MJD $\sim$ 52500 and $\sim$ 53500. These
amplitudes then decreased for 10 orbital cycles until the
Type II outburst was triggered. If the change of outburst amplitudes 
correlated with the 
mass accretion,
then during the decrease 
of these amplitudes mass should be deposited in a disk
around neutron star temporarily. The release of this stored mass may 
ignite the Type II outburst. We report that the optical 
light curve became fainter 
by 0.4 mag during the decrease of amplitude  of the Type I 
outbursts.  The observed H$\alpha$ profiles and  their equivalent widths 
during the decay and after the giant  outburst are consistent with  
previous observations of the system.}
 {}


 \maketitle

\section { Introduction}

 Most of the accretion powered X-ray pulsars have Be type
companions.
Be/X-ray transient binaries exhibit three types of typical behavior:
i) Type II (or giant)
 outbursts which are characterized by high X-ray luminosities
($L_{X} \ge 10^{37}$ erg s$^{-1}$);
ii) Type I (or normal) outbursts which are characterized by
 lower X-ray luminosities ($L_{X} \sim 10^{36}$ erg s$^{-1}$)
 and are generally repeated in each orbital period cycle;
iii) outburst free (quiescence), where the accretion is partially or
 completely halted (Stella et al. 1986; Motch et al. 1991;
 Bildsten et al. 1997).

 The size of the isolated Be star's disk could be resolved from the
variable trend
 of infrared observations. This cannot be measured when the Be star
 is in a binary system since the Be disk is truncated at a resonance
 radius by tidal forces at the orbit of the neutron star
(Okazaki $\&$ Negueruela 2001). In these systems, variations in the mass
loss from the Be star cause the change in the disk density. Therefore
radiation becomes optically thick at infrared wavelengths
(Negueruela et al. 2001; Miroshnichenko et al. 2001).
In this case, infrared magnitudes and the H$\alpha $ equivalent width are
related to the disk density (Negueruela et al. 2001; Miroshnichenko et al.
2001; Wilson et al. 2002).
In most cases,
X-ray outbursts and optical activity in a single outburst do not show
clear correlation. However there are a few exceptions such as those seen in the
outburst of the V635 Cas and X-ray companion 4U 0115+635 (Baykal et al. 2005).

EXO 2030+375 is a 42 s transient accreting X-ray pulsar which was
discovered during a giant outburst in 1985
(Parmar et al. 1989) with a binary orbital period of 46.02 days.
 It has
a B0 Ve spectral type optical companion (Motch $\&$ Janot-Pacheco 1987,
Janot-Pacheco et al. 1988; Coe et al. 1988). The continuous observations of
EXO 2030+375 were performed by Burst and Transient Source Experiment
(BATSE) on the Compton Gamma Ray Observatory from April 1991
until June 2000 (Wilson et al. 2002). The extensive monitoring campaign
is continued by Rossi Timing Explorer (RXTE) (Wilson et al. 2005). These observations
have shown that EXO 2030+375 has undergone an outburst close to the
periastron passage for $\sim$13.5 years.
In 1995, Wilson et al. (2002, 2005) observed
a sudden shift in the orbital phase of EXO 2030+375's outburst from
about 6 days after periastron to about 4 days before periastron.
The outburst phase slowly recovered
to peak about 2.5 days after periastron. They interpreted these phase shifts
in orbital phase as evidence of a global one-armed oscillation propagating
in the Be disk.
 The variations in the shape of the H$\alpha $ profile support
the reconfiguration of the Be disk.
 \begin{figure*}
 \centering
 \includegraphics[clip=true,scale=0.6,angle=270]{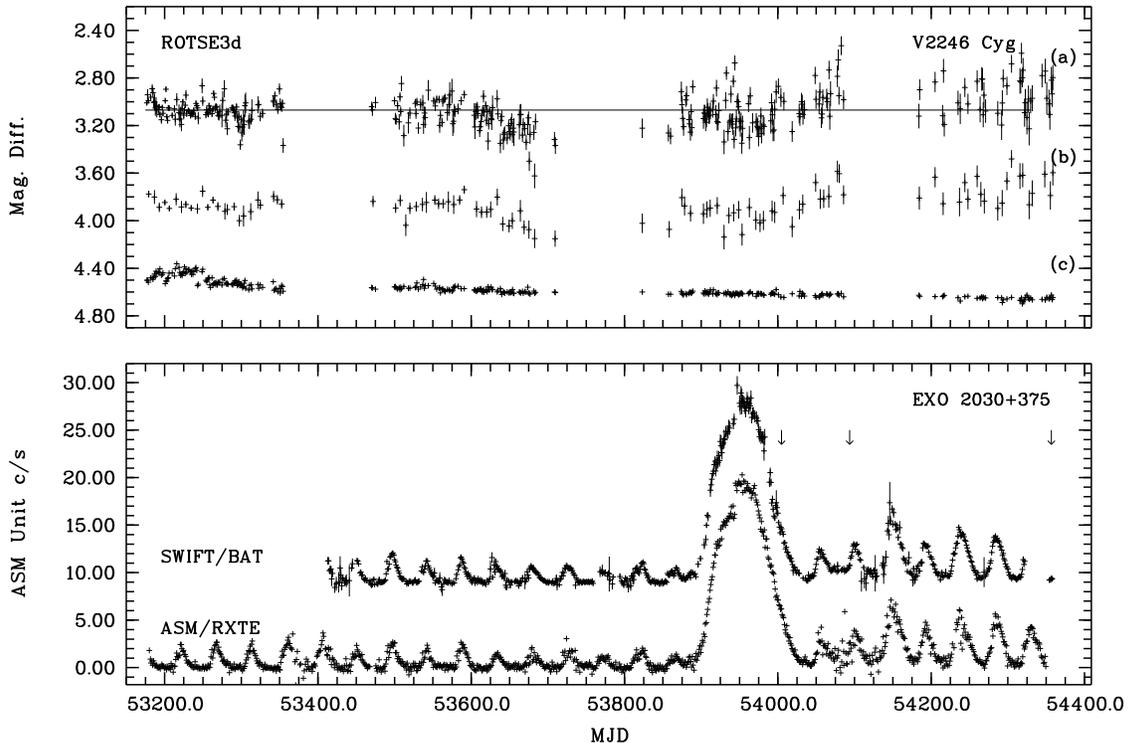}
 \caption{Daily and weekly averages of the differential lightcurve
         of the optical counterpart (\so) to \sx ({\bf a} and {\bf b}),
          and, the mean lightcurve of reference stars ({\bf c}),
          properly offset are shown in the top panel. RXTE/ASM
          lightcurve in 5-12 keV band is shown in the lower panel.
          On the same panel SWIFT/BAT 15-50 keV lightcurve is shown
          with proper offset and scaling.
         }
 \label{fig.1}
 \end{figure*}

RXTE All-Sky Monitor observations of the Be/neutron star
 binary EXO 2030+375 showed
 an increase in flux in mid-June 2006 which indicated that
a new outburst began (Corbet $\&$ Levine 2006, Wilson $\&$ Finger 2006,
Klochkov et al. 2007).  This new outburst stayed for $\sim $ 140 days.
The same  outburst was also observed by SWIFT/BAT. 
EXO 2030+375 was monitored by Robotic Optical Transient Experiment
\footnote{http://www.rotse.net} 
(ROTSE IIId)  and Russian Turkish 1.5 m Telescope
\footnote{http://www.tug.tubitak.gov.tr} (RTT150)
located at Bak{\i}rl{\i}tepe, Antalya, Turkey. In this work we present
the optical and X-ray (RXTE/ASM and SWIFT/BAT) observations of this source.

\section{Observation and data reduction}

\subsection { Optical photometric observations}

 \begin{figure*}
 \centering
 \includegraphics[clip=true,scale=0.5,angle=270]{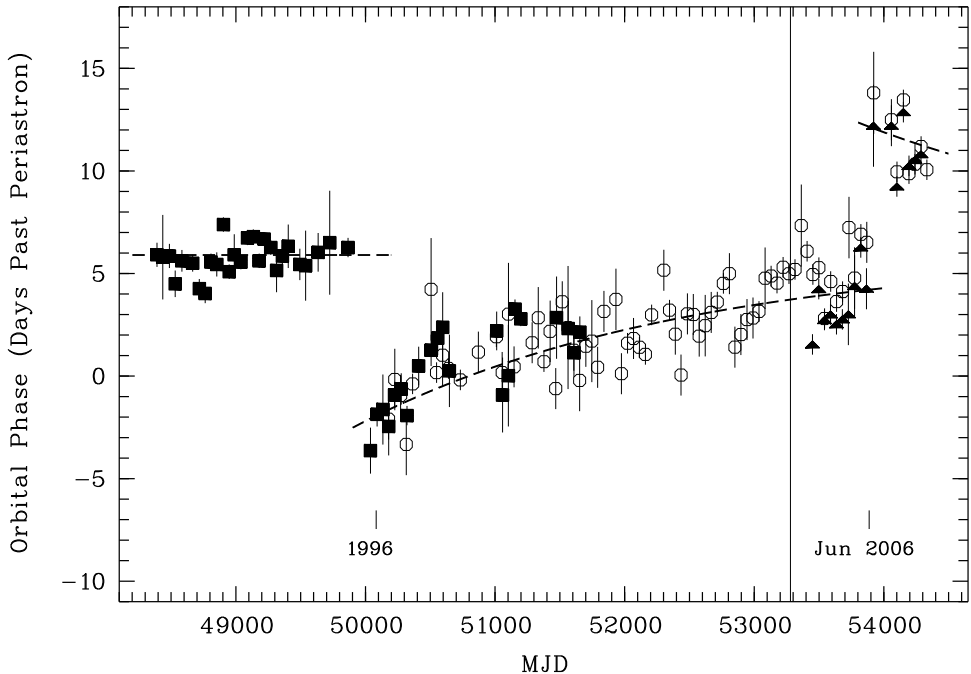}
 \caption{Open circles are Type I outburst arrival times with respect to
 the periastron passage
 extracted from the ASM/RXTE lightcurve with the method described in the text.
 Filled triangles are the arrival times for SWIFT/BAT. The points with
 errors larger than 3 days on the Gaussian centroid are not plotted for clarity.
 Note that the arrival times prior to the vertical line denote the
 recalculated values for the previously published
 orbital phases by Wilson et al. (2002). Squares indicate  BATSE measurements
 taken from Wilson et al. (2002).
 Exponential decay of the major shift around  MJD 50000 with time a scale of
 2500 days is shown. The large shift
 after the June 2006 Type II outburst, is also fitted with the same time scale.
         }
 \label{fig.2}
 \end{figure*}

 The CCD observations of ROTSEIIId were performed from  June 2004 to
September 2007 with a 45 cm
 reflecting telescope. ROTSEIII telescopes operate without filters. They are
equipped with a CCD, 2048 $\times $ 2048 pixel which have the pixel scale
3.3\arcs ~pixel$^{-1}$ for a total field of view 1.85\deg$\times$1.85\deg (Akerlof et
al., 2003).
During the course of the observations, $\sim$1200 CCD frames were collected.
By considering the other scheduled observations
and atmospheric conditions,
1-20 frames with an exposure time of 60 s were taken each night.
All images were automatically dark and flat-field corrected as soon as
they were  exposed. For each corrected image, aperture photometry was 
applied using a 5 pixel (17 arc sec) diameter aperture to obtain the 
instrumental magnitudes. These magnitudes were calibrated by comparing 
all the field stars against the USNO A2.0 R-band catalog.
We adopted three stars as references, whose properties are given
in Table 1. We used their average light curve to reduce long-term
systematic errors. We did not choose stars which had magnitudes 
closer to that of  the target as reference stars, since the magnitude 
determining accuracy decreases for fainter stars in ROTSEIIId measurements 
 (Fig.\ 5 of K{\i}z{\i}lo\u{g}lu et al.\ 2005a). In this way additional 
noise  into the light curve of the target was avoided. We also checked the 
variability of the reference stars. Barycentric correction was made to the 
time of each observation by using JPL DE200 ephemerides. Details of 
the data reduction are described in
K{\i}z{\i}lo\u{g}lu, K{\i}z{\i}lo\u{g}lu $\&$ Baykal (2005a,b).

Figure 1 shows the light curve of daily and weekly averages
of optical counterpart to EXO 2030+375 (\so).
 \begin{table}
 \centering
 \begin{minipage}{140mm}
 \caption{ \so ~and photometric reference stars.}
 \label{table:1}
 \begin{tabular}{@{}ccccc@{}}
 \hline\hline
  Star  & R.A.      & Decl.     &  USNO B1  \\
        & (J2000.0) & (J2000.0) &  R mag    \\
 \hline
 \so    & 20\hr 32\mm 15\fsec22 & +37\deg 38\arcm 15\farcs1 & 17.6  \\
 Star 1 & 20\hr 32\mm 36\fsec49 & +37\deg 36\arcm 19\farcs4 & 13.8  \\
 Star 2 & 20\hr 32\mm 08\fsec32 & +37\deg 40\arcm 26\farcs2 & 13.7  \\
 Star 3 & 20\hr 31\mm 59\fsec63 & +37\deg 37\arcm 17\farcs8 & 13.8  \\
 \hline
 \end{tabular}
 \end{minipage}
 \end{table}
 \begin{table}
 \centering
 \begin{minipage}{140mm}
 \caption{Journal of spectroscopic observations for H$\alpha$ line.}
 \begin{tabular}{@{}lclcccc@{}}
 \hline
 Date & MJD & Grism
         \footnote{Nominal dispersion for G15 is 3 \AA ~pixel$^{-1}$. \\
           Hence 3 pixel resolution at H$\alpha$ line is  $\sim$9 \AA.}
             & Exp. & FWHM & EW \\
      &     &       & (sec) & (\AA) & (\AA)  \\
 \hline
 2006 Sep 26& 54004 & G15  & 3600 & 22.5$\pm$7.6 &~~8.7$\pm$1.1 \\
 2006 Dec 24& 54093 & G15  & 3600 & 18.9$\pm$5.5 &~~8.9$\pm$1.2 \\
 2007 Sep 13& 54356 & G15  & 5400 & 16.3$\pm$4.3 &10.2$\pm$1.3 \\
 \hline
 \end{tabular}
 \end{minipage}
 \end{table}

\subsection { Optical spectroscopic observations}

The spectroscopic observations were performed on September 26, and 
December 24, 2006 and September 13, 2007 with the Russian-Turkish
1.5 m Telescope (RTT150) using medium resolution spectrometer TFOSC
(T\"UB\.ITAK Faint Object Spectrometer and Camera). We used grism G15
(spectral range 3200-9125 \AA) with the nominal dispersion of
$\sim$3.1 \AA ~pixel$^{-1}$.
Table 2 gives the measurements of the
equivalent width (EW) of the H$\alpha$ line for each observing run
together with the values of full width at half maximum (FWHM).

The reduction and analysis of spectrum was made 
 using the package Longslit context of MIDAS. The FWHM
and EW values were obtained from Gauss fit models to
 the emission features.

 \begin{table*}
 \centering
 \caption{Type I outburst epochs from ASM/RXTE and SWIFT observations}
 \begin{center}
 \begin{tabular}{cc cc cc cc}
 \hline
 RXTE/ASM &&&&&&& \\
 \hline
 Orbit & Epoch &Orbit & Epoch &Orbit & Epoch &Orbit & Epoch \\
 Number & (MJD) & Number & (MJD) &Number & (MJD) &Number & (MJD) \\
 \hline
 0  &50178.0 $\pm$ 1.0 & 23 &51231.2 $\pm$ 3.5 & 45 &52259.0 $\pm$ 4.5 & 67 &53268.4 $\pm$ 0.5 \\
 1  &50225.9 $\pm$ 1.5 & 24 &51286.2 $\pm$ 1.0 & 46 &52302.1 $\pm$ 1.0 & 68 &53314.6 $\pm$ 0.5 \\
 2  &50271.2 $\pm$ 1.0 & 25 &51333.4 $\pm$ 1.5 & 47 &52346.2 $\pm$ 0.5 & 69 &53362.8 $\pm$ 2.0 \\
 3  &50314.8 $\pm$ 1.5 & 26 &51377.3 $\pm$ 0.5 & 48 &52391.1 $\pm$ 1.0 & 70 &53407.5 $\pm$ 0.5 \\
 4  &50363.8 $\pm$ 0.5 & 27 &51424.8 $\pm$ 1.5 & 49 &52435.1 $\pm$ 1.0 & 71 &53452.4 $\pm$ 0.5 \\
 5  &50410.8 $\pm$ 5.0 & 28 &51468.0 $\pm$ 1.0 & 50 &52484.1 $\pm$ 1.0 & 72 &53498.8 $\pm$ 0.5 \\
 6  &50457.3 $\pm$ 5.0 & 29 &51518.3 $\pm$ 1.0 & 51 &52530.1 $\pm$ 1.0 & 73 &53542.3 $\pm$ 0.5 \\
 7  &50506.4 $\pm$ 2.5 & 30 &51563.0 $\pm$ 3.0 & 52 &52575.0 $\pm$ 1.0 & 74 &53590.1 $\pm$ 0.5 \\
 8  &50548.4 $\pm$ 0.5 & 31 &51607.9 $\pm$ 1.0 & 53 &52621.6 $\pm$ 1.5 & 75 &53635.2 $\pm$ 0.5 \\
 9  &50595.3 $\pm$ 1.5 & 32 &51652.5 $\pm$ 1.5 & 54 &52668.2 $\pm$ 1.0 & 76 &53681.7 $\pm$ 0.5 \\
 10 &50640.6 $\pm$ 0.5 & 33 &51700.2 $\pm$ 1.0 & 55 &52714.8 $\pm$ 0.5 & 77 &53730.8 $\pm$ 1.5 \\
 11 &50686.8 $\pm$ 5.0 & 34 &51746.4 $\pm$ 2.0 & 56 &52761.7 $\pm$ 0.5 & 78 &53774.4 $\pm$ 1.5 \\
 12 &50732.1 $\pm$ 0.5 & 35 &51791.2 $\pm$ 1.0 & 57 &52808.2 $\pm$ 1.0 & 79 &53822.6 $\pm$ 0.5 \\
 14 &50825.7 $\pm$ 5.0 & 36 &51839.9 $\pm$ 1.0 & 58 &52850.6 $\pm$ 1.0 & 80 &53868.2 $\pm$ 1.0 \\
 15 &50871.5 $\pm$ 1.0 & 37 &51879.7 $\pm$ 5.0 & 59 &52897.3 $\pm$ 1.0 & 81 &53921.5 $\pm$ 2.0 \\
 16 &50917.3 $\pm$ 5.0 & 38 &51932.6 $\pm$ 1.5 & 60 &52944.0 $\pm$ 1.0 & 84 &54058.2 $\pm$ 1.0 \\
 17 &50962.1 $\pm$ 5.0 & 39 &51975.0 $\pm$ 1.0 & 61 &52990.1 $\pm$ 1.0 & 85 &54101.7 $\pm$ 0.5 \\
 18 &51010.3 $\pm$ 0.5 & 40 &52022.5 $\pm$ 0.5 & 62 &53036.4 $\pm$ 0.5 & 86 &54151.2 $\pm$ 0.5 \\
 19 &51054.6 $\pm$ 1.0 & 41 &52068.7 $\pm$ 1.0 & 63 &53084.1 $\pm$ 1.5 & 87 &54193.7 $\pm$ 0.5 \\
 20 &51103.5 $\pm$ 2.5 & 42 &52114.3 $\pm$ 0.5 & 64 &53130.2 $\pm$ 0.5 & 88 &54240.1 $\pm$ 0.5 \\
 21 &51146.9 $\pm$ 1.0 & 43 &52160.0 $\pm$ 0.5 & 65 &53175.9 $\pm$ 0.5 & 89 &54287.0 $\pm$ 0.5 \\
 22 &51188.6 $\pm$ 3.5 & 44 &52207.9 $\pm$ 0.5 & 66 &53222.7 $\pm$ 0.5 & 90 &54331.9 $\pm$ 0.5
 \\
 \hline
 SWIFT/BAT &&&&&&& \\
 \hline
 71 &53449.0 $\pm$ 0.5 & 76  &53680.3 $\pm$ 0.5 & 81 &53919.9 $\pm$ 2.0 & 88 &54240.3 $\pm$ 0.5 \\
 72 &53497.7 $\pm$ 0.5 & 77  &53726.6 $\pm$ 1.5 & 84 &54057.9 $\pm$ 1.0 & 89 &54286.6 $\pm$ 0.5 \\
 73 &53542.2 $\pm$ 0.5 & 78  &53774.0 $\pm$ 1.5 & 85 &54101.0 $\pm$ 0.5 &    & \\
 74 &53588.5 $\pm$ 0.5 & 79  &53821.9 $\pm$ 0.5 & 86 &54150.6 $\pm$ 0.5 &    & \\
 75 &53634.9 $\pm$ 0.5 & 80  &53865.9 $\pm$ 1.0 & 87 &54194.0 $\pm$ 0.5 &    &
 \end{tabular}
 \end{center}
 \end{table*}

\subsection  {
RXTE/ASM and SWIFT/BAT observations}
The All Sky Monitor (ASM) on board the Rossi X-ray Timing Explorer (RXTE) satellite
 consists of three wide-angle Scanning Shadow Cameras (SSCs). These cameras
are mounted on a rotating drive assembly, which covers $\sim $ 70$\%$ of the sky every
1.5h
 (Levine et al. 1996). ASM light curves can be found in the public archive
in different energy bands (1.3-3.0, 3.0-5.0, 5.0-12.0 keV). In Fig. 1,
we present the  ASM  light curve in 5-12 keV.

 The burst Alert Telescope (BAT) on board SWIFT satellite
provides 15-150 keV monitoring observations primarily to determine
the positions of gamma-ray bursts. The light curves of several targets are
available in the public archive of SWIFT/BAT
\footnote{http:swift.gsfc.nasa.gov/docs/swift/results/}. 15-50 keV BAT light curve of
EXO 2030+375 is also presented in Fig. 1.

\subsection {Optical and X-ray variability}

 We searched the short term variability in the optical light curve by applying
Scargle and Clean algorithms (Scargle 1982, Roberts, Lehar $\&$ Dreher 1987).
We did not detect any significant periodicity in the optical light curve.

EXO 2030+375 showed a Type I outburst for each orbital cycle 46.02 days.
 Detailed studies of these outbursts were performed by Wilson et al.
(2002, 2005). In these studies, it was found that prior to $\sim $1996,
outbursts were about 6 days after periastron passages. Then suddenly
outbursts shifted to about 4 days before periastron. The orbital phase
of the outburst was rapidly recovered 2.5 days after periastron, and
continued to change gradually. Finally, as of 
November 2004, the orbital phase of the outburst peak reached $\sim $ 5 days
after periastron (Wilson et al. 2005).

In order to keep track, the phases of the Type I outbursts were estimated  
using archival observations of RXTE/ASM and SWIFT/BAT.
We fitted a Gaussian model to   the Type I outburst light curves of 
RXTE/ASM (5-12 keV) and SWIFT/BAT (15-150 keV) observations. 
We centered the Gaussian profile on the  approximately predicted
Type I outburst passage time. In the fitting, the center value, the amplitude
and the width of Gaussian were taken as free
parameters. We then determined the peak time of the outburst.

In Fig. 2, we present the observed minus calculated values
of outburst epochs
 ($T_{outburst}-n<P_{orbit}>-T_{periastron}$) relative to the
constant orbital period ($<P_{orbit}>=46.0214$ days).
The epoch of the periastron passage from the pulse timing solution was 
obtained  by Wilson et al. (2002) as MJD 50547.22. To obtain the closest 
periastron passage to the first X-ray outburst,     
we subtracted 8 orbital cycles from this epoch
($T_{periastron}=50547.22-8*46.02=$MJD~$50179.06$).
It is clear that our results are consistent with previously
published curves (Wilson et al. 2002, 2005; Camero et al. 2005) and
indicate that the outburst delay varies around 4-7 days prior to the giant
outburst. In Table 3, we display the outburst arrival times and orbital 
cycle numbers n. 

 \begin{figure}
 \centering
 \centering
 \includegraphics[clip=true,scale=0.33,angle=270]{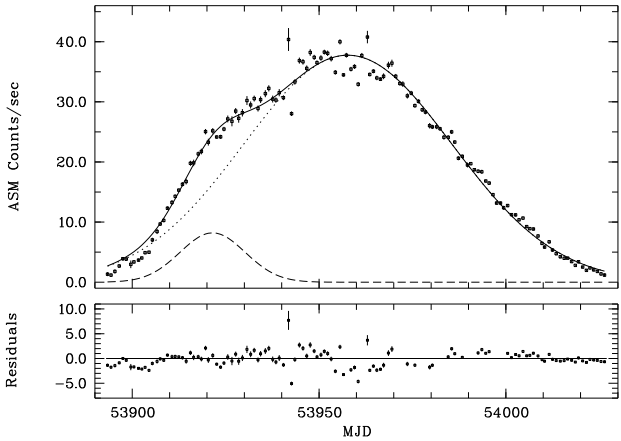}
 \caption{The best fit to the giant outburst by two Gaussian model (above)
   and its residuals (below).  The Gaussian model describing the
 Type I outburst is centered at MJD 53921 and has width 8.35 days while the
 other Gaussian is centered at MJD 53958 with width of 31.8 days.
         }
 \label{fig.3}
 \end{figure}

 \begin{figure}
 \centering
 \includegraphics[clip=true,scale=0.33,angle=270]{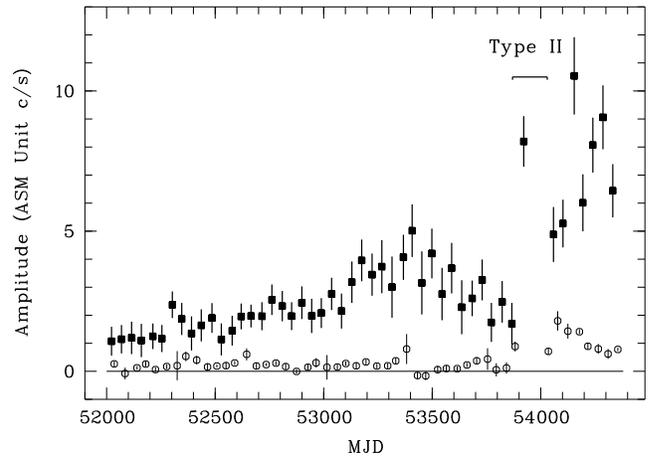}
 \caption{Filled squares are the best fit amplitudes of Type I outbursts
          for the RXTE/ASM light curve. Note the gradual increase of
          the amplitudes prior to
          rapid decrease after MJD 53500 followed by onset of the giant
          outburst around MJD 53870.
          Open circles are the baseline estimates of the count rates between
          Type I outbursts. Note also the decaying behavior of the baseline
          after MJD 54050.
         }
 \label{fig.4}
 \end{figure}

We modeled a Type II outburst using a simple two component Gaussian model.
 Figure 3 shows the two component Gaussian fit to the Type II outburst.
The best fit is obtained by two Gaussian curves:
the main peak being centered at MJD 53957.8 with a width of 31.8 days and 
a secondary peak centered at MJD 53921.5 with a width of 8.35 days.
 The residuals to the fit are also shown in the same figure. 
The peak time for the secondary Gaussian corresponds 
to a shift of 13.8 days after the periastron passage. 
This secondary peak can be interpreted as a Type I outburst
in the sense that its peak time and amplitude are in agreement with the
results obtained for other Type I outburst peaks 
observed after the giant Type II outburst,
as seen in Fig. 2 and Fig. 4. Since Type I outbursts occurred periodically 
for a long time before the giant outburst, we expected this secondary peak 
to be a continuation of periodic behavior during the Type II outburst. 

 We estimated the amplitudes of Type I outbursts 
 from Gaussian fits for each outburst.
We also estimated the baseline count rates by fitting a 
parabola to the RXTE/ASM light curve between the successive Type I outbursts.
In Fig. 4, we present the time evolution of amplitudes of the Type I outburst
and baseline estimates.
The amplitudes of the Type I outbursts are fairly stable around 1 c/s before
MJD 52500.  Then a gradual increase of the amplitudes up to MJD 53500 and
a rapid decrease before the onset of the giant Type II outbursts is seen.
After the Type II outburst, the amplitudes of the Type I outbursts show an 
increase and  EXO 2030+375 brightens in Type I X-ray outburst.

The baseline count rate estimates of RXTE/ASM observations are 
$\sim$0.2 c/s before the Type II outburst.
After the giant outburst, the baseline count rates decrease  
starting from $\sim$2 c/s. 
In order to see the decreasing trend more clearly, we fitted an 
exponential  model to the baseline count rate estimates.  
The fitted exponential decay time is 155$\pm$21 days. 
The same type of behavior is present for SWIFT observations 
with a decay time of 134$\pm$13 days.

\section{Discussion}

 The infrared and spectroscopic observations of Wilson et al. (2002)
have  shown that the density of the disk and
 the brightness of IR observations are correlated.
Their observations also showed that the equivalent width of H$\alpha$ increased
with the density of the disk. Based on the study of IR and X-ray observations,
they concluded that the Be disk was truncated at a 4:1 resonance radius.
 According to their findings,
 in 1993 a major structural change  occurred in the
Be star's circumstellar disk. After about January 1993, the JHK magnitudes
began to decline and dropped to a minimum near August 1996 and 
both the  IR magnitudes and the H$\alpha$ equivalent widths
 indicated that the density of disk decreased.  
 \begin{figure}
 \centering
 \includegraphics[clip=true,scale=0.3,angle=270]{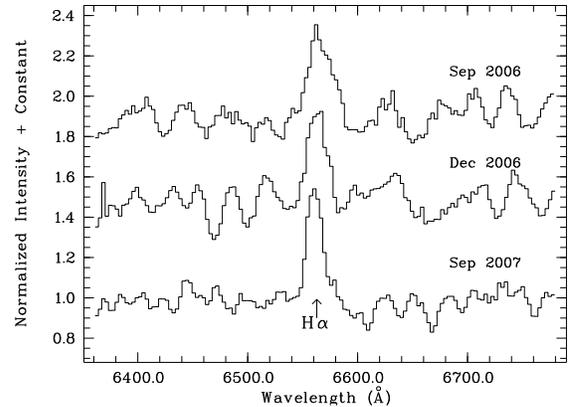}
 \caption{H$\alpha$ profiles observed in moderate resolution spectrum of the
  counterpart to EXO 2030+375 (V2246Cyg) taken on Sep 26, and Dec 24, 2006
  and Sep 13, 2007 (MJD 54004.8,  54093.7 and 54356.8, with grism 15.
  3 pixel resolution at the H$\alpha$ line is
  $\sim$9 \AA ). }
 \label{fig.5}
 \end{figure}

  ROTSE observations are excellent for continuous monitoring
 of optical counterpart (\so) ~to EXO 2030+375.
Fig. 1 shows that the optical counterpart to EXO 2030+375 becomes fainter
after MJD 53600 (during the decrease of amplitudes of Type I outbursts).
This faintness in the magnitudes probably shows an increase in the
density of the disk which may result during the sudden discrete ejection
of material by the Be star to its circumstellar disk. This material
may obscure light coming from the photosphere of Be star.
The change of about 0.4
magnitude in the optical light is similar to the change
that we observed in the optical light of V635 Cas before the Type II
outburst of 2004 (Baykal et al. 2005).
After MJD 53900 (during and after the Type II outburst) the optical
observations did not show variabilities in the magnitudes above
one sigma confidence level. We also did not detect any optical
variability in this system during the Type I outbursts as Norton et al.
(1994) noted.

H$\alpha$ ($\lambda$6563 \AA) profile is a good  indicator for the presence
of circumstellar disk around Be stars and its morphology. The observed
H$\alpha$ profiles and their EW during  the decay of 
and after the giant Type II outburst are presented in Fig. 5 and Table 2.
 The observed profiles are single peaked. However, the data of September 2006
seem more complex than a single Gaussian which
 shows an asymmetric profile with a red shoulder.
This indicates a perturbed density distribution in the Be disk
during the Type II outburst (Negueruela et al. 1998).
The asymmetric profile of H$\alpha$ line begins to disappear 3 months later. 
 Due to low flux in  the observed profiles 
and poor resolution, it is difficult to say more about the shape 
of profiles.
However, the measured EW is consistent with the  
H$\alpha$ EW values of  Reig et al. (1998) and Wilson et al. (2002) 
 during the X-ray active period after MJD 50000.

Formation of a transient disk 
around neutron star during Type II outburst can be predicted considering 
the baseline estimates of the count rates.
 After the Type II outburst, the baseline count rate is about 
2 c/s. It then decreases to a 
level similar to pre-outburst rates with 
a decay time of 155$\pm$21 days suggesting that a transient disk 
is formed around the neutron star during the giant outburst. 
The transient disk seems to disappear in about 5 months.  
Klochkov et al. (2007) have predicted a sharp change in the spin
frequency of the pulsar in this system.  They attributed this change to 
the formation  of an accretion disk around the neutron star.  
Motch et al. (1991) makes a similar suggestion about the formation of 
a transient accretion disk during the giant outburst of the Be/X-ray 
system A0535+26 considering the steep dependency of the specific angular 
momentum of the accreted material with envelope expansion velocity during 
the Type II outburst.

Type I X-ray outbursts peak
 $\sim $6 days after periastron passage prior to the Type II outburst.
After the Type II outburst, Type I outburst peaks show a sudden shift to 
$\sim $13 days after periastron passage.
Negueruela et al. (1998) proposed that there is a relation between 
density waves and Type II outbursts.  The perturbation in the density waves 
would cover the whole radial  structure of  the disk with a typical time 
scale of a few days to years. They observed evidence of density 
perturbation coinciding with Type II outbursts for Be binary 
systems of 4U 0115+634 and A0535+26. We expect similar behavior 
for EXO 2030+375.

Negueruela et al. (1998), also suggested that  Type I outbursts may be observed
where the interaction of neutron star occurred with the regions of enhanced
density at which these waves are generated. 
The precession period of density waves for EXO 2030+375 was suggested to be 
15$\pm$3 years by Wilson, Fabregat $\&$ Coburn (2005). Recent observations
imply that this precession period is interrupted during the Type II outburst.
A new precession period could be set up but at this point the period
is not known. Future monitoring of this source should yield a 
better understanding of this source.

\begin{acknowledgements}
This project utilizes data obtained by the Robotic Optical Transient Search
Experiment.  ROTSE is a collaboration of Lawrence Livermore National Lab,
Los Alamos National Lab and the University of Michigan
(http://www.rotse.net).
We thank the Turkish National Observatory of T\"UB\.ITAK
for running the optical facilities.
We acknowledge support from T\"UB\.ITAK, The Scientific and Technological
Research Council
of Turkey,  through project 106T040.
We also acknowledge the RXTE/ASM team for the X-ray monitoring data.
Swift/BAT transient monitor results provided by the Swift/BAT team.
\end{acknowledgements}

\end{document}